# Remote Sensing D/H Ratios in Methane Ice: Temperature-Dependent Absorption Coefficients of CH$_3$D in Methane Ice and in Nitrogen Ice

W.M. Grundy[a], S.J. Morrison[b], M.J. Bovyn[c], S.C. Tegler[c], and D.M. Cornelison[d]

a. Lowell Observatory, 1400 W. Mars Hill Rd., Flagstaff AZ 86001.
b. Cornell University, Dept. of Astronomy, Ithaca NY 14853; 2010 summer REU student at Northern Arizona University.
c. Northern Arizona University, Dept. of Physics and Astronomy, Box 6010 Flagstaff AZ 86011.
d. Missouri State University, Dept. of Physics, Astronomy, and Materials Science, 901 S. National Ave., Springfield MO 65897; formerly at Northern Arizona University Dept. of Physics and Astronomy.

**ABSTRACT**

The existence of strong absorption bands of singly deuterated methane (CH$_3$D) at wavelengths where normal methane (CH$_4$) absorbs comparatively weakly could enable remote measurement of D/H ratios in methane ice on outer solar system bodies. We performed laboratory transmission spectroscopy experiments, recording spectra at wavelengths from 1 to 6 µm to study CH$_3$D bands at 2.47, 2.87, and 4.56 µm, wavelengths where ordinary methane absorption is weak. We report temperature-dependent absorption coefficients of these bands when the CH$_3$D is diluted in CH$_4$ ice and also when it is dissolved in N$_2$ ice, and describe how these absorption coefficients can be combined with data from the literature to simulate arbitrary D/H ratio absorption coefficients for CH$_4$ ice and for CH$_4$ in N$_2$ ice. We anticipate these results motivating new telescopic observations to measure D/H ratios in CH$_4$ ice on Triton, Pluto, Eris, and Makemake.

## 1. Introduction

Deuterium (D or $^2$H) is a stable isotope of hydrogen (H) with one neutron in addition to the usual proton. Most of the deuterium in the universe is thought to have originated in the big bang (e.g., Epstein et al. 1976) and it is consumed by nuclear reactions in stars, so the universe's total inventory declines over time. In cold clouds of dust and gas, deuterium is strongly fractionated into the dust particles (e.g., Herbst 2003). The deuterium enrichment of the dust component of the Sun's natal cloud provides opportunities to trace the evolution of that material during solar system formation. Where presolar solids were vaporized they would return their deuterium to the nebular gas, but to the extent that these particles retain their deuterium-rich phases, bodies accreting from that solid material would be correspondingly enriched. Processing of presolar dust particles is expected to have varied across the nebula, with more presolar grains surviving in colder nebular environments further from the Sun (e.g., Fegley 1999). In the coldest, outermost parts of the nebula, deuterium could also fractionate into nebular dust particles via the same mechanisms responsible for its enrichment in interstellar dust (Fegley 1999), leading to even greater enrichment of the solids in those regions.

Evidence for contrasting D/H ratios among solar system bodies is well established. The terrestrial D/H ratio of $1.5 \times 10^{-4}$ (Robert et al. 2000) is considerably higher than the initial bulk solar system value, estimated to be $2.1 \times 10^{-5}$ from observations of the atmospheres of Jupiter and Saturn (Lellouch et al. 2001). $H_2O$ gas observed in the comae of four long period comets that may have formed in the Jupiter and Saturn forming region prior to being scattered into the Öort cloud (e.g., Duncan 2008) shows higher D/H ratios around $\sim 3 \times 10^{-4}$ (e.g., Meier and Owen 1999; Altwegg and Bockelée-Morvan 2003; Villanueva et al. 2009). Extreme heterogeneity of D/H in carbonaceous chondrites, and even within individual interplanetary dust particles (where localized regions with D/H ratios as high as $\sim 8 \times 10^{-3}$ have been found), is indicative of highly variable processing of presolar grains as well as complex mixing of nebular solids (e.g., Messenger 2002; Busemann et al. 2006; Duprat et al. 2010). Uranus and Neptune show enrichment as well, attributed to the substantial quantities of outer nebular solids they accreted in addition to nebular gas (e.g., Lutz et al. 1990; Encrenaz 2005).

The diverse populations of small, icy, outer solar system bodies that accreted from solids in peripheral parts of the protoplanetary nebula present a wealth of opportunities to sample additional nebular environments; measuring D/H ratios among these populations could be extremely valuable (e.g., Horner et al. 2007, 2008). However, subsequent processes could have significantly modified isotopic ratios on these bodies. Volatile loss would have played a particularly important role in their evolution (e.g., Schaller and Brown 2007), elevating D/H ratios in their remaining volatile ices. Additional processes that could continue to influence D/H ratios on the surfaces of these bodies include seasonal volatile transport cycles, photolysis and radiolysis, and ion-molecule chemistry (e.g., Brown and Cruikshank 1997). Measuring D/H ratios could be valuable for studying those processes, as well. Some outer solar system small body populations can be sampled by observing comets. Ecliptic comets are thought to originate somewhere in the Kuiper belt (e.g., Duncan 2008), but we are not aware of any reported D/H ratios for ecliptic comets. It would be even better to measure D/H for outer solar system bodies *in situ*, since opportunities may not exist to observe comets from some of the more interesting small body reservoirs, such as irregular satellites of giant planets, Trojans of Jupiter and Neptune, and detached Kuiper belt objects such as Sedna. Also, by the time an object has become an ecliptic comet, it is hard to tell exactly where in the Kuiper belt it originated.

Unfortunately, no current observational technique is capable of measuring D/H ratios in small outer solar system bodies. However, Pluto, Eris, and Makemake, three planet-sized objects in the Kuiper belt, all have near-infrared reflectance spectra dominated by strong vibrational absorptions of $CH_4$ ice, offering the possibility of measuring D/H ratios in their surface ice. To these we can add Neptune's large, retrograde satellite Triton, thought to have been captured from the Kuiper belt. When one hydrogen atom in a $CH_4$ molecule is replaced with deuterium, the result is singly deuterated methane $CH_3D$. The altered mass balance and symmetry of this molecule produces a distinct pattern of infrared vibrational absorption bands relative to ordinary $CH_4$. At wavelengths where ordinary $CH_4$ absorbs least, photons can traverse the longest path lengths within that ice before being scattered out of the surface and potentially observed. A strong $CH_3D$ band falling in one of these window regions would be particularly sensitive for measuring D/H ratios in $CH_4$ ice by reflectance spectroscopy. $CH_3D$ has been studied extensively in the gas phase, and also condensed in liquid argon and in low temperature, cubic α $N_2$ ice (e.g., Nelander 1985; Blunt et al. 1996; Nikitin et al. 2006). These studies show that among the many near-infrared $CH_3D$ absorptions, some do occur in regions lacking strong $CH_4$ absorptions, making measurement of D/H ratios in $CH_4$ ice via remote observation seem possible. We could not find published absorption coefficients for $CH_3D$ in $CH_4$ ice at temperatures appropriate for the surfaces of Triton, Pluto, Eris, and Makemake, so we undertook a series of laboratory transmission spectroscopy experiments to obtain the necessary data. We also studied $CH_3D$ diluted in the higher temperature, hexagonal β phase of $N_2$ ice because $CH_4$ diluted within that phase is known to exist on Triton and Pluto (e.g., Quirico and Schmitt 1997; Quirico et al. 1999; Douté et al. 1999).

## 2. Experimental Procedures

The experiments reported in this paper were done in a new laboratory ice facility located in the Department of Physics and Astronomy of Northern Arizona University. Ice samples were crystallized within an enclosed cell (Fig. 1) fitted with windows to allow a spectrometer beam to pass through the ice. Detailed descriptions of this facility and our sample measurement procedures have been published by Tegler et al. (2010). Subsequent to that paper, we implemented several crucial improvements.

The primary change was the addition of an infrared MCT (mercury cadmium telluride) type A detector, cooled with liquid nitrogen. This detector is sensitive from below 1 µm to beyond 10 µm, although our use of sapphire windows on the sample cell limits us to operating in the ≤ 6 µm range. Still, this is a substantial extension over our wavelength coverage using only the silicon detector described by Tegler et al. (2010). Operating at infrared wavelengths necessitated several additional changes. To overcome infrared absorptions intrinsic to our system, we replaced our glass lenses with off-axis aluminum paraboloid mirrors (see Fig. 2), and replaced the glass windows on the vacuum enclosure with sapphire windows. Absorption by water vapor and $CO_2$ in the ambient room air was minimized by bagging as much of the optical path as we could and purging the bagged volume with air from which these two species had been removed by a Whatman purge gas generator (unfortunately, it was not possible to purge the entire optical path and still be able to easily monitor the cell contents by eye). For best sensitivity at shorter wavelengths ($\lambda < 2.5$ µm) we used a quartz halogen lamp and a quartz beamsplitter. At longer wavelengths ($\lambda > 2.5$ µm), an infrared glowbar light source and a KBr (potassium bromide) beamsplitter gave better results. In practice, many experiments were done twice, recording a complete temperature series using each configuration. The substantial region of overlapping wavelengths between the two (roughly 1.7 to 3.4 µm) provided a useful check on consistency of results.

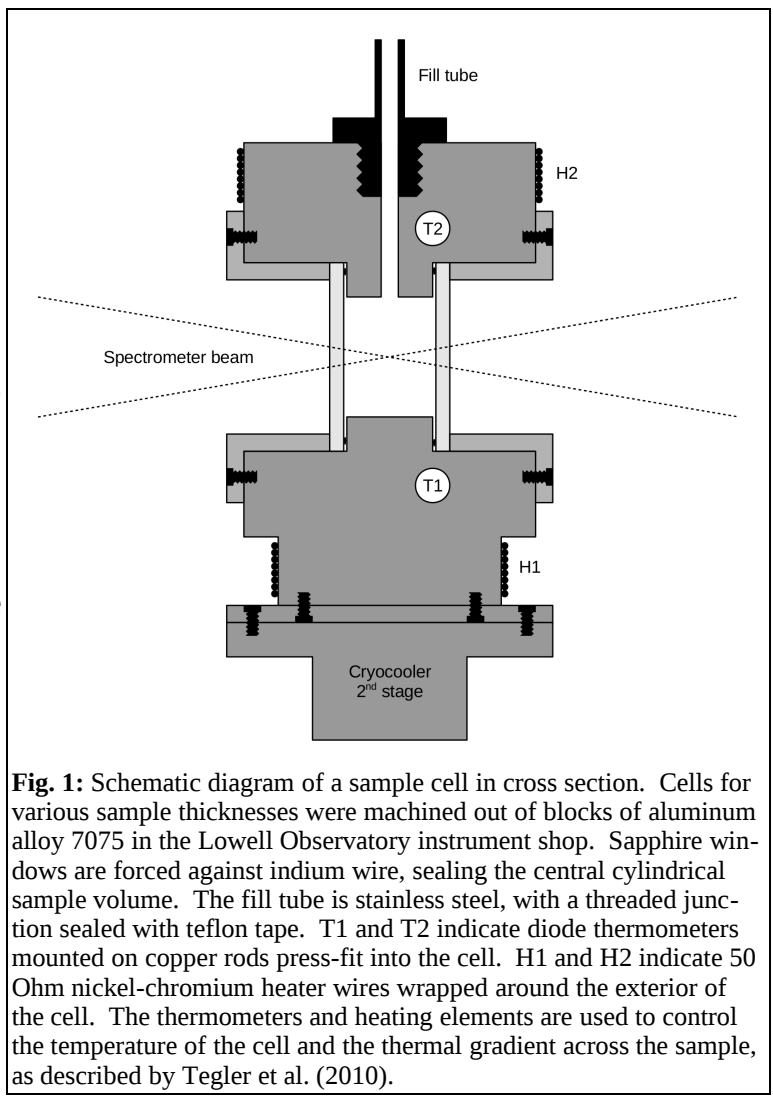

**Fig. 1:** Schematic diagram of a sample cell in cross section. Cells for various sample thicknesses were machined out of blocks of aluminum alloy 7075 in the Lowell Observatory instrument shop. Sapphire windows are forced against indium wire, sealing the central cylindrical sample volume. The fill tube is stainless steel, with a threaded junction sealed with teflon tape. T1 and T2 indicate diode thermometers mounted on copper rods press-fit into the cell. H1 and H2 indicate 50 Ohm nickel-chromium heater wires wrapped around the exterior of the cell. The thermometers and heating elements are used to control the temperature of the cell and the thermal gradient across the sample, as described by Tegler et al. (2010).

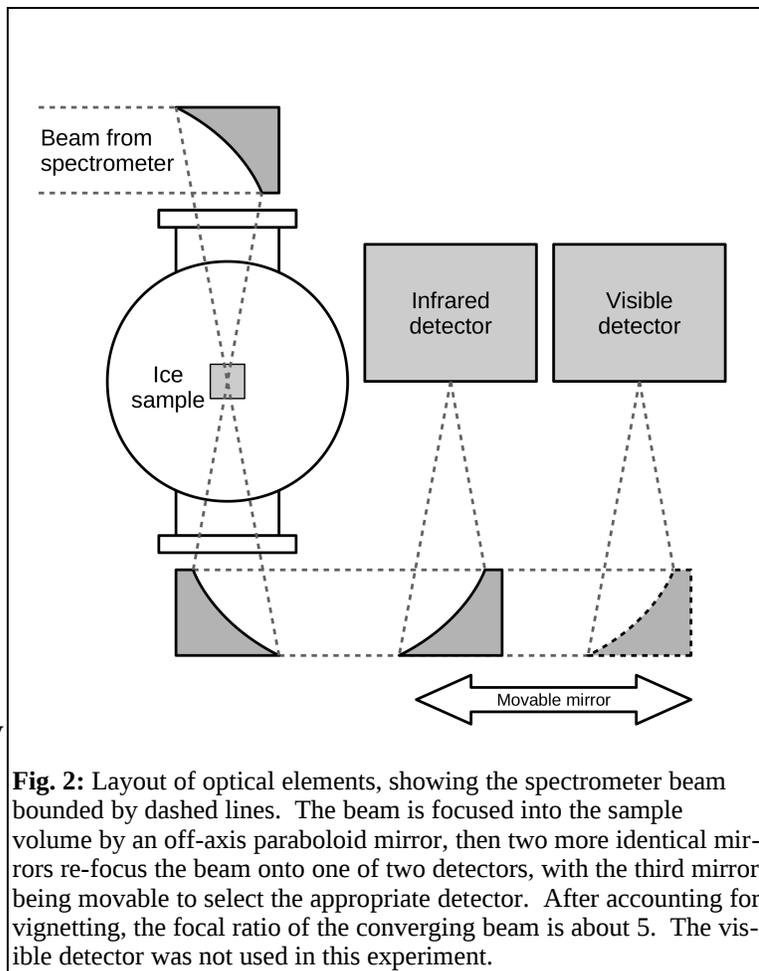

**Fig. 2:** Layout of optical elements, showing the spectrometer beam bounded by dashed lines. The beam is focused into the sample volume by an off-axis paraboloid mirror, then two more identical mirrors re-focus the beam onto one of two detectors, with the third mirror being movable to select the appropriate detector. After accounting for vignetting, the focal ratio of the converging beam is about 5. The visible detector was not used in this experiment.

    To facilitate preparation of mixed samples, we also added a mixing volume of approximately 2 liters, monitored by a pair of Baratron capacitance manometers covering the range from $10^{-3}$ to $10^{+3}$ torr. Samples were mixed at room temperature in this volume using $CH_4$ and $N_2$ gases supplied in high pressure cylinders by Airgas Specialty Gases. These were connected through separate pressure regulators and valves to the mixing volume. The reported purities were ≥ 99.99% for the $CH_4$ and ≥ 99.9% for the $N_2$. To these gases, we could add $CH_3D$ purchased in a lecture bottle from Sigma Aldrich (product number 490237, with a reported purity of ≥ 98%). Unlike in water, proton-deuteron exchange is negligible in methane at room temperature and below, so we did not need to consider the formation of other isotopomers such as $CH_2D_2$, $CHD_3$, etc. (Sigma Aldrich Stable Isotope Department, personal communication 2010). After mixing gases to the desired composition at room temperature in the mixing volume, we opened a valve to allow the gas to flow into the empty, cold cell, condensing it as a liquid. We froze this liquid by reducing its temperature, maintaining a vertical thermal gradient of about 2 K across the 15 mm diameter sample by means of heaters (see Fig. 1), so that it would freeze from the bottom upward, with the location of the freezing front being controlled by the cell temperature. Each new mixture requires some experimentation to find appropriate cooling rates, but rates for freezing samples were typically in the range of −0.01 to −0.1 K/min. It was sometimes necessary

to freeze an initial polycrystalline mass and then melt almost all of it to obtain a small seed crystal before re-freezing slowly, in order to obtain an optical-quality ice sample. After a suitable sample had been frozen, we removed the thermal gradient by smoothly shifting temperature-control heating from the upper to the lower heating element over a period of 10 to 20 minutes. Using only the bottom heater for temperature control resulted in a thermal gradient across the sample of just a few tenths of a Kelvin.

Spectra were recorded with a Nicolet Nexus 670 Fourier transform infrared (FTIR) spectrometer at a sampling interval of 0.24 cm$^{-1}$, resulting in a spectral resolution of 0.6 cm$^{-1}$ (measured full width at half maximum of unresolved lines). The spectrometer beam was focused to a few mm spot inside the cell. We typically averaged over 100 spectral scans to improve the signal/noise ratio. After a sample spectrum had been recorded, we would ramp to a new temperature, at rates typically in the range of 0.1 to 0.5 K/min. We recorded spectra through our ice samples at every multiple of 10 K between 40 K and the host ice melting points (90.7 K for CH$_4$ ice, 63.1 K for N$_2$ ice).

Before each ice sample was prepared we also recorded spectra through the empty, cold cell, and did the same after each sample was eliminated. Filled-cell spectra were divided by empty-cell spectra to remove gross effects of lamp emission, detector sensitivity, and absorptions by the windows and air, resulting in approximate transmission spectra $T(\lambda)$. These spectra are affected by subtle, spurious slopes from a variety of sources. Wavelength-dependent refractive index contrasts exist between ice and cell windows leading to slightly different transmission through the ice-window interface than through the vacuum-window interface. The room temperature refractive index $n(\lambda)$ of sapphire decreases gradually with wavelength from about 1.76 to 1.59 from 1.0 to 5.5 µm (Malitson et al. 1958; Gervais 1991) but methane ice shows almost no wavelength dependence in its refractive index (except at wavelengths near 3.3 µm where absorption is so strong that we measure no transmission whatsoever; Pearl et al. 1991). If we knew the temperature dependent $n(\lambda)$ of both sapphire and ice, and the ice-window interfaces were solely responsible for these slopes, we could easily correct for the effect. But as the ice and the surrounding cell contract on cooling, each with their own distinct temperature-dependent coefficients of thermal expansion, the sample is stressed and can fracture or pull away from the windows, opening additional ice-vacuum interfaces that can produce wavelength-dependent scattering that varies with temperature, with thermal history, and with location within the sample. In addition to imparting slopes, these effects lead to a decline in overall transmission, by as much as a factor of three in an ice sample cooled relatively rapidly from 90 to 40 K. Slow drifts in lamp filament temperature or detector sensitivity over the course of experiments lasting multiple days can also contribute spurious slopes. Slopes arising from any combination of the above factors were removed by fitting a line or low-order polynomial to continuum regions adjacent to absorption bands to be quantified and dividing by this function to "straighten out" the continuum. The continuum-corrected transmission spectra were then converted to Lambert absorption coefficient spectra $\alpha(\lambda)$ via the Beer-Lambert absorption law, rearranged as $\alpha(\lambda) = -\ln(T(\lambda))/d$, where $d$ is the path length through the cell ($d$ = 5.4 ± 0.1 mm for all experiments reported in this paper).

## 2.1 CH$_3$D diluted in CH$_4$

We added a small amount of CH$_3$D to CH$_4$ to produce a deuterium-enriched sample that remained dominated by normal CH$_4$ absorptions. Addition of 0.5% CH$_3$D (D/H ratio of $1.25 \times 10^{-3}$) produced ice having sufficiently strong CH$_3$D absorption to be easily measurable, but not so strong as to be saturated in transmission though our 5.4 mm optical path length. We were not able to directly measure the D/H ratio in our ice sample, so we assumed zero fractionation between the gas phase and the liquid condensed from it, and again zero fractionation between the liquid and the solid crystallized from it. Many ordinary $CH_4$ bands were saturated in this sample, but this was not a problem because we were only interested in wavelengths where $CH_4$ absorbs weakly. Comparing transmission spectra of ordinary $CH_4$ ice with spectra of the enriched ice mixture, three regions between 1 and 6 µm meet our criterion of having strong $CH_3D$ absorption features coinciding with weak $CH_4$ absorption, necessary to detect minute quantities of $CH_3D$ against a background of much more abundant ordinary $CH_4$ ice. Transmission spectra through 40 K ice samples are shown in Fig. 3.

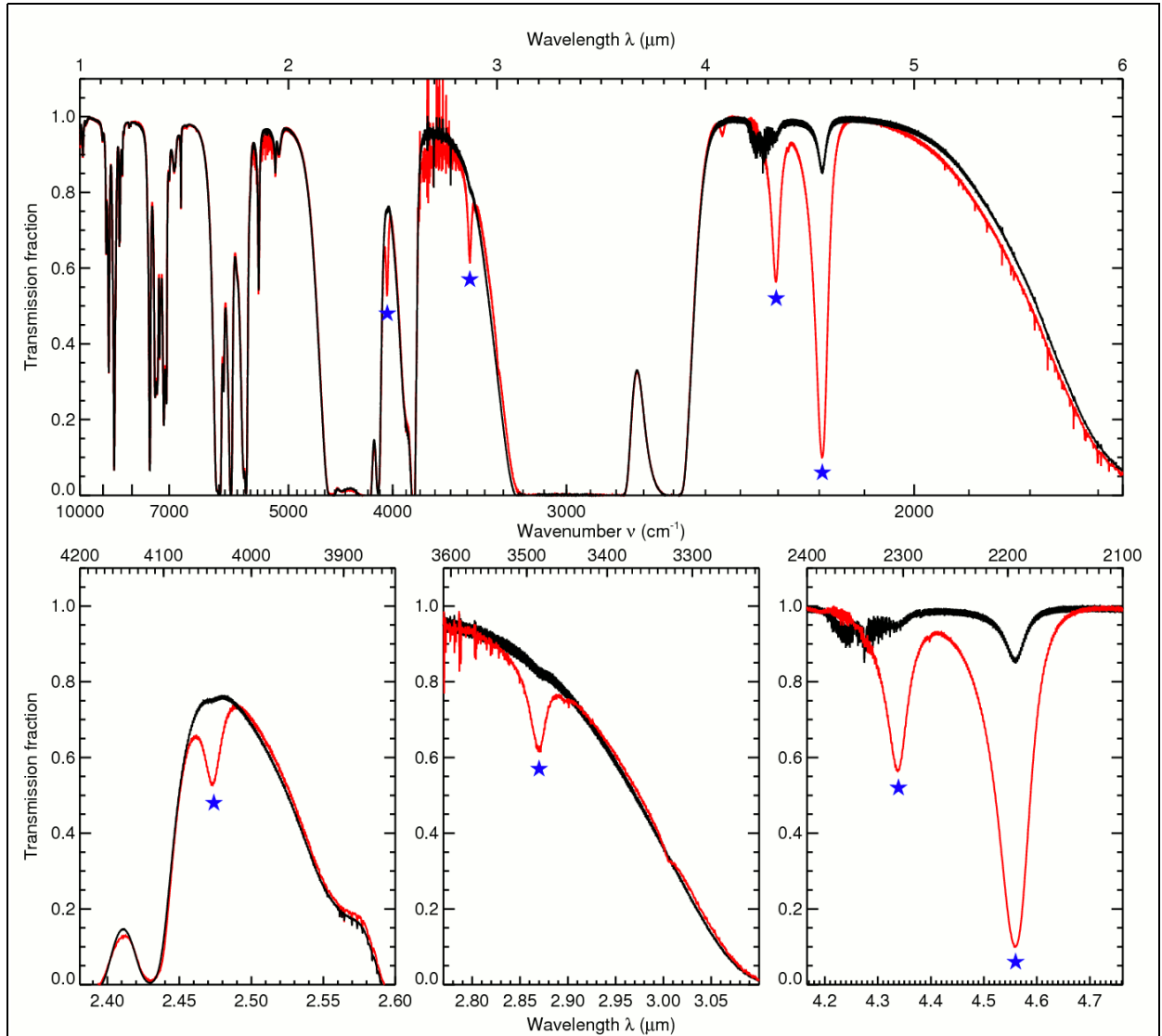

**Fig. 3:** Transmission spectra of normal $CH_4$ ice (black curves) and of $CH_4$ ice enriched with 0.5% $CH_3D$ (red curves), both at 40 K. Potentially useful $CH_3D$ bands are marked with blue stars, and shown in greater detail in the lower panels. These three spectral regions where deuterium-enriched methane exhibits absorptions but normal $CH_4$ shows high transmission are of particular interest for remote spectroscopic measurement of D/H ratios in $CH_4$ ice. Note that at the wavelength of each $CH_3D$ absorption band, the ordinary methane also shows a weak feature, since some $CH_3D$ is present in the ordinary methane. Narrow, spiky features are residual absorptions from ambient air, indicating wavelengths unobservable from ground-based telescopes.

The lower right panel of Fig. 3 shows a region in which Grundy et al. (2002) had tentat-ively identified two bands at 4.34 and 4.56 µm as being caused by $CH_3D$. Our new spectra of ordinary $CH_4$ (black curve) and of $CH_4$ enriched in $CH_3D$ (red curve), clearly confirm $CH_3D$ as the cause of these two bands. Even the $CH_4$ spectrum shows some absorption at these wavelengths, because there is a small amount of $CH_3D$ present in ordinary laboratory-grade $CH_4$. The noisy region around 4.20 to 4.35 µm results from imperfect cancellation of strong $CO_2$ absorptions from the ambient air. That spectral region is unobservable from ground-based telescopes, but could be observed from space-based platforms. The 4.56 µm band occurs in the *M*-band atmospheric window, and could potentially be observed from the ground. The lower middle panel shows a $CH_3D$ absorption band at 2.87 µm, on the shoulder of the strong $v_3$ fundamental of $CH_4$ centered at 3.3 µm. The 2.87 µm band roughly coincides with the short wavelength limit of the *L*-band atmospheric window. The lower left panel shows a $CH_3D$ absorption band at 2.47 µm, in a region of relatively weak absorption between adjacent strong $CH_4$ bands. The 2.47 µm band falls within the *K*-band atmospheric window (albeit near the long wavelength limit). Another comparably strong $CH_3D$ band is raising the continuum at the left of this band, but strong absorption by ordinary methane at those wavelengths prevents us from computing accurate absorption coefficients, and would, in any case, limit its value for remote sensing of D/H ratios in $CH_4$ ice. Both the 2.47 and 2.87 µm $CH_3D$ bands could potentially be observed from a high altitude, ground-based site.

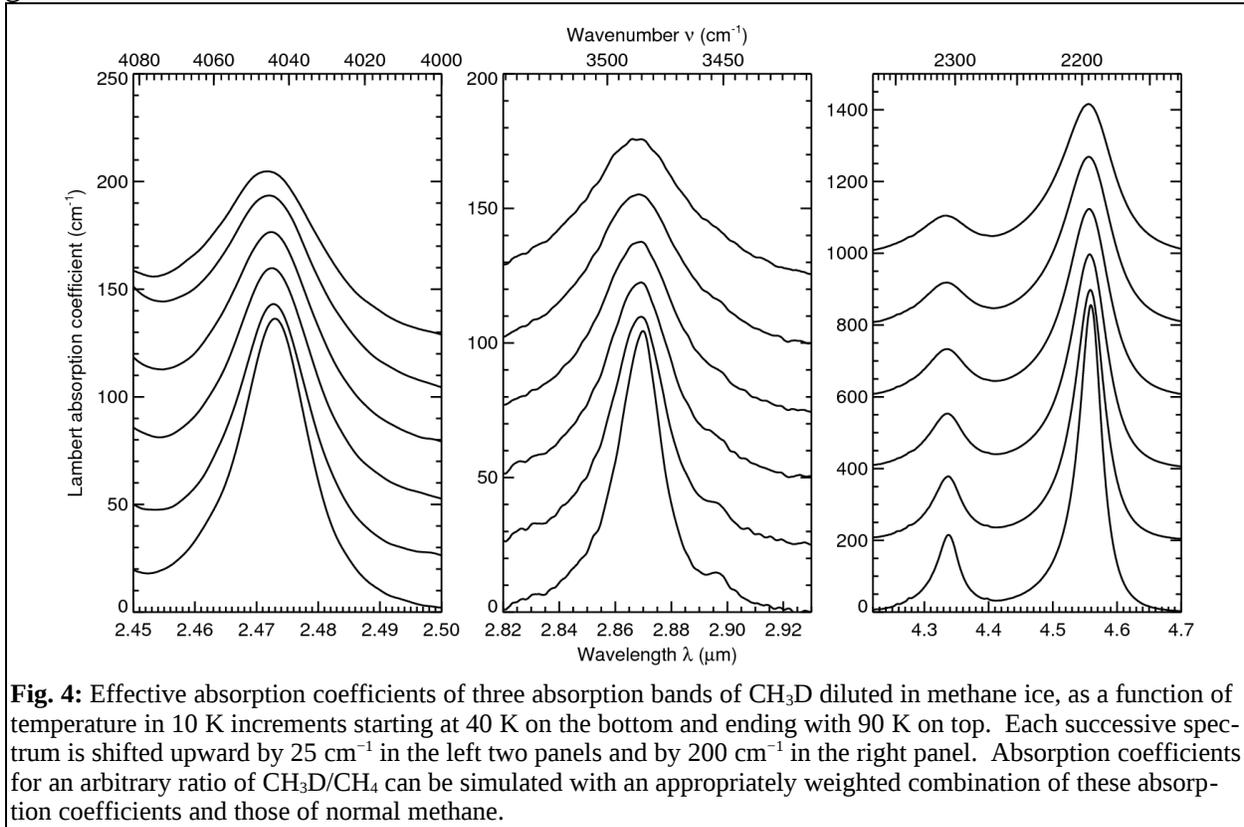

**Fig. 4:** Effective absorption coefficients of three absorption bands of $CH_3D$ diluted in methane ice, as a function of temperature in 10 K increments starting at 40 K on the bottom and ending with 90 K on top. Each successive spectrum is shifted upward by 25 cm$^{-1}$ in the left two panels and by 200 cm$^{-1}$ in the right panel. Absorption coefficients for an arbitrary ratio of $CH_3D/CH_4$ can be simulated with an appropriately weighted combination of these absorption coefficients and those of normal methane.

In each of these three wavelength regions, we subtracted the ordinary methane absorption coefficients (computed from the black curves in Fig. 3) to isolate the contribution of the $CH_3D$. The resulting $CH_3D$ ice absorption coefficients were divided by the 0.5% concentration of the $CH_3D$ (assuming the composition was unchanged by condensation and freezing) to obtain effective absorption coefficients for $CH_3D$ in methane as if it were pure $CH_3D$. These values, shown in Fig. 4, can be combined with ordinary $CH_4$ ice absorption coefficients, scaled by their relative abundances, to approximate absorption coefficients for arbitrary $CH_3D/CH_4$ mixing ratios, as will be discussed in Section 3.

For all of these bands, some effect of $CH_3D$ absorption can also be seen in the $CH_4$ ice spectrum, indicating that the ordinary $CH_4$ used in our experiments has a non-zero fraction of $CH_3D$ in it. The effect of this small quantity of $CH_3D$ needs to be subtracted off to properly distinguish between the absorptions of $CH_3D$ and $CH_4$. Assuming that all of the 4.56 µm absorption band can be attributed to $CH_3D$, we can use our new $CH_3D$ absorption coefficients to determine how much $CH_3D$ was present in our $CH_4$ gas, by finding the amount of $CH_3D$ absorption that must be subtracted to eliminate any trace of the observed band. The result is a $CH_3D/CH_4$ ratio of $(3.2 \pm 0.1) \times 10^{-4}$, corresponding to a D/H ratio of $(8.0 \pm 0.3) \times 10^{-5}$. Applying the same procedure to the $CH_4$ absorption coefficients published by Grundy et al. (2002), using methane from Air Liquide Corporation in France, we obtain very similar $CH_3D/CH_4$ and D/H values of $(3.3 \pm 0.1) \times 10^{-4}$ and $(8.3 \pm 0.3) \times 10^{-5}$, respectively. Naturally occurring terrestrial methane has D/H ratios in the $1.1 \times 10^{-4}$ to $1.4 \times 10^{-4}$ range (Schoell 1980). This is depleted relative to the D/H ratio of $\sim 1.5 \times 10^{-4}$ in terrestrial water, but less depleted than we estimate for both laboratory methane samples, leading us to speculate that the purification process used to remove higher molecular mass hydrocarbon contaminants also removes some $CH_3D$. Alternatively, we could have assumed that the ordinary methane in both labs produced ice with a typical terrestrial methane D/H ratio of $1.25 \times 10^{-4}$, in which case our $CH_3D$-enriched ice sample must have had a D/H ratio a factor of 1.6 higher than the $CH_3D$-enriched gas from which it was made. Our reported absorption coefficients would then be too high by this same factor of 1.6. This uncertainty about the composition of the ice sample substantially limits the precision of our derived absorption coefficients.

From our temperature series we were able to study the temperature-dependent behavior of the $CH_3D$ absorption bands. At lower temperatures, they become narrower. The same behavior occurs in normal $CH_4$ ice (Grundy et al. 2002). The narrowing effect reveals the presence of additional weak side bands, such as one at about 2.897 µm. Also at lower temperatures, all three of the strong $CH_3D$ bands show a subtle shift to longer wavelengths. In frequency units, this shift averages 2.2 cm$^{-1}$ over the temperature interval from 90 to 40 K. Such a thermal shift has not, to our knowledge, been reported in ordinary $CH_4$. It could be an indication of a subtle, temperature-dependent vibrational coupling between $CH_3D$ molecules and neighboring $CH_4$ molecules.

We note that numerous additional $CH_3D$ bands exist, such as at 1.56, 1.89, 1.92, 1.94, 2.24, 2.40, and 3.01 µm. Some of these other absorptions have been used previously for measuring $CH_3D/CH_4$ ratios in gaseous methane in outer solar system atmospheres (e.g., Lutz et al. 1983; de Bergh et al. 1986, 1988, 1990), and some of these bands can be seen in our data as well. But compared with the three bands we focus on, these other $CH_3D$ absorptions are either much weaker or they coincide with stronger ordinary $CH_4$ ice absorptions. To measure meaningful D/H ratios using these other bands would require telescopic observations with much higher signal precision.

## 2.2 CH₃D diluted in N₂

Much of the methane ice observed in the outer solar system exhibits bands shifted to slightly shorter wavelengths, indicative of dilution in another ice. On Pluto and Triton, that ice has been identified as the hexagonal β phase of $N_2$ ice by direct observation of the 2.15 µm absorption band of $N_2$ ice (e.g., Cruikshank et al. 1984; Owen et al. 1993; Grundy et al. 1993). On Eris and Makemake, smaller $CH_4$ shifts have also been observed and tentatively attributed to dilution in $N_2$ ice (e.g., Licandro et al. 2006a,b; Dumas et al. 2007; Abernathy et al. 2009; Merlin et al. 2009; but see also Tegler et al. 2010 who point out that argon ice can also shift $CH_4$ bands, much as nitrogen ice does). Where the concentration of methane exceeds its solubility limit in $N_2$ ice, methane-rich and nitrogen-rich phases will coexist (e.g., Prokhvatilov and Yantsevich 1983; Lunine and Stevenson 1985), a situation that has already been seen in spectra of Pluto and Eris (Douté et al. 1999; Tegler et al. 2010).

To investigate possible effects of dilution in $N_2$ ice on the absorption bands of $CH_3D$, we performed an experiment with $CH_3D$ diluted in the hexagonal β phase of $N_2$ ice. $N_2$ ice melts at 63.1 K, so this experiment spanned a smaller range of temperatures than described in the previous section. Uncertainty over the $CH_3D$ concentration in the ice presented even more of a challenge with this experiment. The $CH_3D/N_2$ ratio mixed in the gas phase in the mixing volume could not be expected to remain unchanged in the ice, for two reasons. First, methane is much less volatile than nitrogen. On condensing the $N_2$-dominated gas into the cell as a liquid at about 65 K, some of the deuterated methane could have condensed as frost somewhere in the inlet tube rather than making it into the cell. Second, compositional gradients appear on freezing, as a result of the separation between liquidus and solidus curves of the binary phase diagram of nitrogen and methane (e.g., Prokhvatilov and Yantsevich 1983; unfortunately, the two curves are difficult to distinguish in their figure). Quirico and Schmitt (1997) assumed that the integrated absorptions of $CH_4$ bands remain unchanged on dilution in $N_2$ ice in order to estimate the composition of their samples. Making the same assumption for $CH_3D$, we used the integrated absorption of the 4.56 µm $CH_3D$ band from the previous section to estimate the $CH_3D$ fraction in our mixed $CH_3D+N_2$ ice sample as 0.0024 ± 0.0002 (about a factor of two below its gas phase abundance of 0.005).

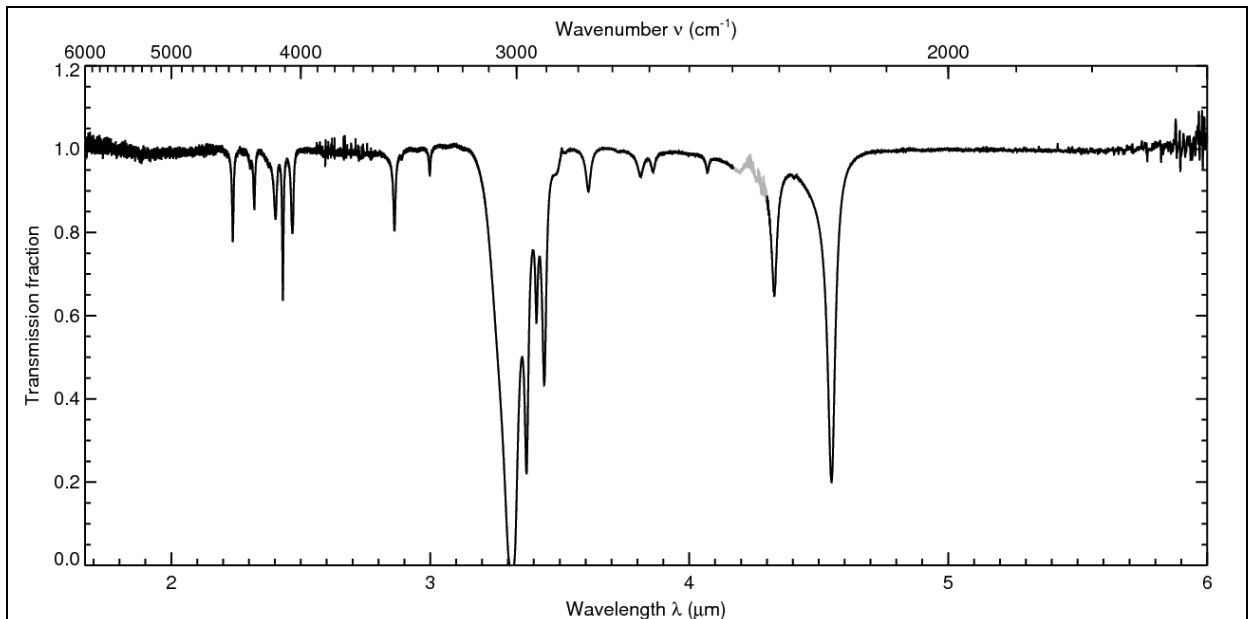

**Fig. 5:** Transmission spectrum of $CH_3D$ diluted in β $N_2$ ice at 40 K, divided by a transmission spectrum of $N_2$ ice at the same temperature to remove the broad $N_2$ ice absorption around 4.1 to 4.4 µm. This spectrum shows $CH_3D$ ice bands listed in Table 1, including a number of bands that are masked by strong $CH_4$ absorptions when ordinary $CH_4$ ice is present. The gray area is particularly strongly affected by atmospheric $CO_2$ absorption.

As before, we subtracted the absorption coefficients of the host $N_2$ ice, studied in an otherwise identical separate experiment in our laboratory, to isolate the $CH_3D$ absorptions. Unlike in $CH_4$ ice, saturated bands of the host ice were not a problem in this experiment and we were able to measure $CH_3D$ bands over a broader range of wavelengths. A transmission spectrum of $CH_3D$ diluted in $N_2$ ice at 40 K is shown in Fig. 5, revealing numerous absorption features listed in Table 1. Absorption bands of $N_2$-diluted $CH_3D$ are compared with their $CH_4$-diluted counterparts in Fig. 6. On dilution in $N_2$ ice, we observed the $CH_3D$ bands shift to shorter wavelengths and become narrower. This behavior is both qualitatively and quantitatively very similar to that of ordinary $CH_4$ bands diluted in $N_2$ ice (Quirico and Schmitt 1997).

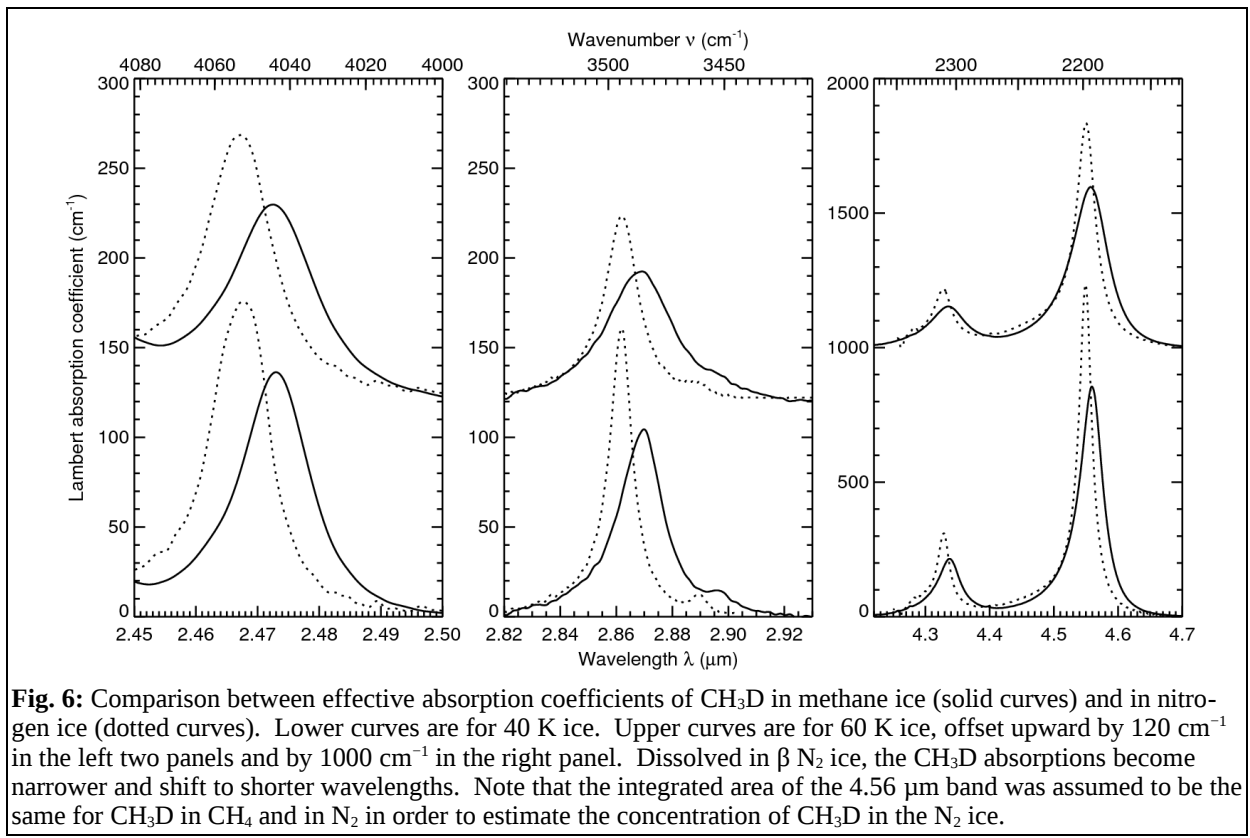

**Fig. 6:** Comparison between effective absorption coefficients of $CH_3D$ in methane ice (solid curves) and in nitrogen ice (dotted curves). Lower curves are for 40 K ice. Upper curves are for 60 K ice, offset upward by 120 cm$^{-1}$ in the left two panels and by 1000 cm$^{-1}$ in the right panel. Dissolved in β $N_2$ ice, the $CH_3D$ absorptions become narrower and shift to shorter wavelengths. Note that the integrated area of the 4.56 μm band was assumed to be the same for $CH_3D$ in $CH_4$ and in $N_2$ in order to estimate the concentration of $CH_3D$ in the $N_2$ ice.

**Table 1**

**Locations of observed CH$_3$D absorption bands**

| Band assignment | CH$_3$D in CH$_4$ | | CH$_3$D in N$_2$ | |
| --- | --- | --- | --- | --- |
| | λ (μm) | ν (cm$^{-1}$) | λ (μm) | ν (cm$^{-1}$) |
| ν$_4$ + 2ν$_6$ | 1.89 | 5290 | 1.88 | 5310 |
| ν$_4$ + ν$_5$ | - | - | 2.24 | 4470 |
| 2ν$_2$ | - | - | 2.32 | 4340 |
| ν$_4$ + ν$_6$ | - | - | 2.40 | 4160 |
| ν$_1$ + ν$_6$ | - | - | 2.43 | 4110 |
| 2ν$_3$ + ν$_5$ | 2.47 | 4040 | 2.47 | 4050 |
| 3ν$_6$ and ν$_2$ + ν$_3$ | 2.87 | 3470 | 2.86 | 3490 |
| ν$_2$ + ν$_6$ | 3.01 | 3330 | 3.00 | 3340 |
| ν$_4$ | - | - | saturated | saturated |
| ν$_1$ | - | - | 3.37 | 2970 |
| 2ν$_5$ (E) | - | - | 3.41 | 2930 |
| 2ν$_5$ (A$_1$) | - | - | 3.44 | 2910 |
| ν$_3$ + ν$_5$ | - | - | 3.61 | 2770 |
| ν$_5$ + ν$_6$ | - | - | 3.81 | 2620 |
| 2ν$_3$ | - | - | 3.86 | 2590 |
| ν$_3$ + ν$_6$ | 4.08 | 2450 | 4.07 | 2460 |
| 2ν$_6$ | 4.34 | 2310 | 4.33 | 2310 |
| ν$_2$ | 4.56 | 2190 | 4.55 | 2200 |

Table note: Band assignments are from Blunt et al. (1996) and Nikitin et al. (2006). Dashes indicate bands that could not be measured in CH$_4$ ice due to the presence of strong CH$_4$ absorptions.

## 3. Models

Using the absorption coefficients reported here in conjunction with data from the literature, it is possible to simulate absorption coefficients $\alpha_r$ for CH$_4$ ice with an arbitrary D/H ratio $r$, assuming the deuterium is in the form of CH$_3$D and not more highly deuterated species, and ignoring possible concentration-dependent changes in vibrational coupling with neighboring ordinary CH$_4$ molecules. Subject to these assumptions, absorption coefficients can be simulated both for CH$_4$ ice, and for CH$_4$ highly diluted in N$_2$. For CH$_4$ ice, the first step is to compute synthetic deuterium-free CH$_4$ absorption coefficients $\alpha_{CH_4}$ by subtracting the CH$_3$D contribution $\alpha_{CH_3D}$ from the Grundy et al. (2002) absorption coefficients $\alpha_{G'02}$, now that we have estimated the CH$_3$D/CH$_4$ ratio in that sample to have been $3.3 \times 10^{-4}$ (in Section 2.1, assuming no appreciable isotopic fractionation between gas, liquid, and solid phases),

$$\alpha_{CH_4} = 1.00033\,\alpha_{G'02} - 0.00033\,\alpha_{CH_3D}. \tag{1}$$

Next, convert the desired D/H ratio $r$ to a $CH_3D$ fraction $f$, accounting for the numbers of hydrogen atoms in each type of molecule,

$$f = \frac{4r}{1+r}. \tag{2}$$

Finally, use $f$ to weight the separate spectral contributions of the synthetic deuterium-free $CH_4$ and of $CH_3D$ to the simulated absorption coefficients $\alpha_r$,

$$\alpha_r = (1-f)\alpha_{CH_4} + f\alpha_{CH_3D}. \tag{3}$$

For $CH_4$ diluted in $N_2$ ice, the process is the same, except that the starting point is the nitrogen-diluted $CH_4$ absorption coefficients of Quirico and Schmitt (1997) which should share the same $CH_3D/CH_4$ ratio of $3.3 \times 10^{-4}$, since those samples were produced approximately contemporaneously from the same high pressure $CH_4$ gas bottle (again assuming no isotopic fractionation of $CH_4$ on preparation of $N_2$-diluted ice samples in either laboratory).

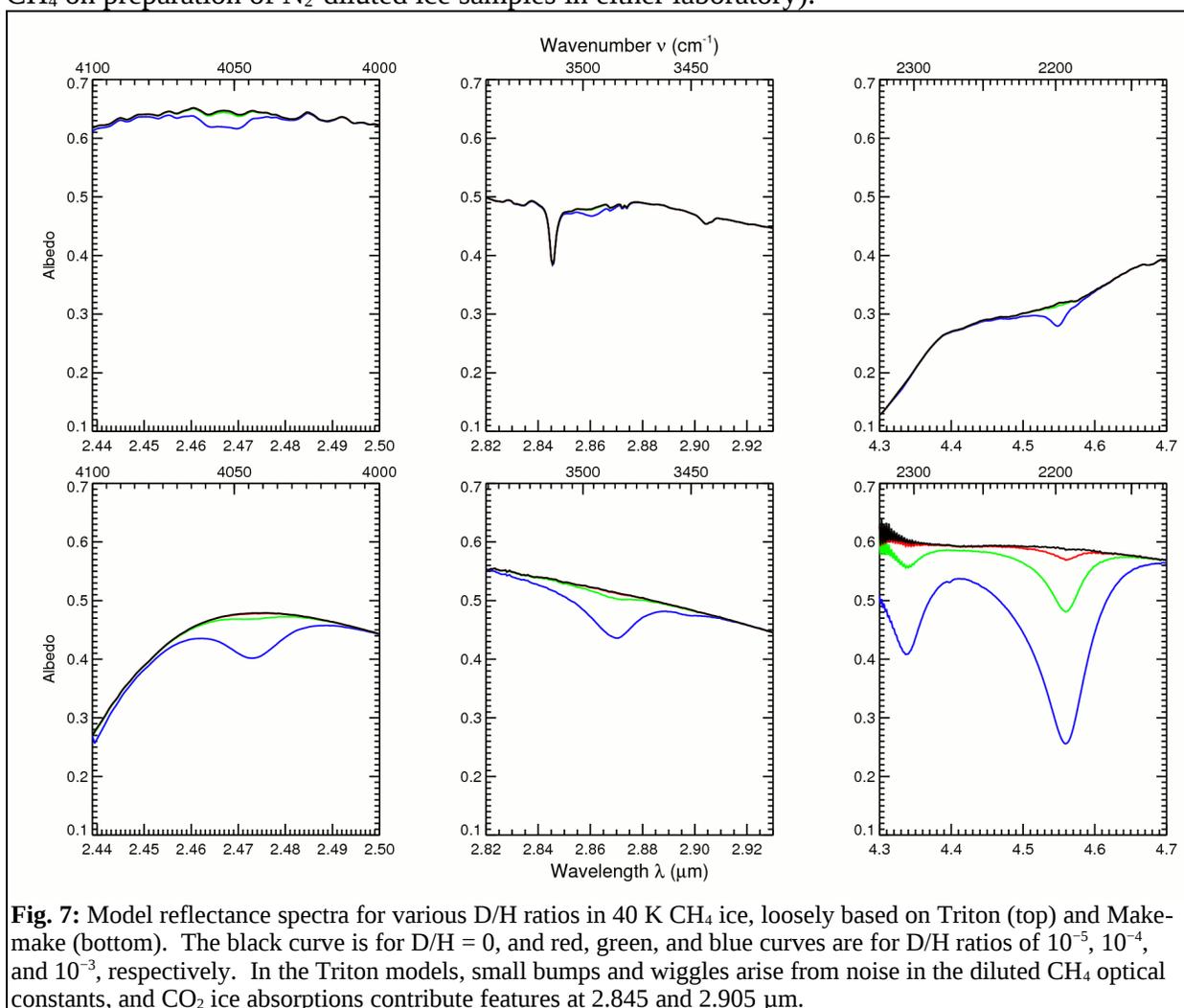

**Fig. 7:** Model reflectance spectra for various D/H ratios in 40 K $CH_4$ ice, loosely based on Triton (top) and Makemake (bottom). The black curve is for D/H = 0, and red, green, and blue curves are for D/H ratios of $10^{-5}$, $10^{-4}$, and $10^{-3}$, respectively. In the Triton models, small bumps and wiggles arise from noise in the diluted $CH_4$ optical constants, and $CO_2$ ice absorptions contribute features at 2.845 and 2.905 µm.

The synthetic absorption coefficients $\alpha_r$ can be input to a radiative transfer model to see how $CH_4$ ice with a particular D/H ratio would appear in reflectance. Better yet, such a model can be numerically inverted to infer the D/H ratio of the ice from observations of $CH_3D$ absorption bands in spectra of outer solar system bodies with strong $CH_4$ absorptions, such as Pluto, Triton, Eris, and Makemake, subject to the additional assumptions inherent in such models. For illustration purposes, we used a Hapke model (e.g., Hapke 1993) to compute example reflectance spectra, accounting for multiple scattering in a compositionally homogeneous, particulate surface. From the depths of their $CH_4$ absorptions, we selected Triton and Makemake as bounding cases. Triton has the shallowest $CH_4$ bands of the four, and Makemake has the deepest. Additionally, Triton's $CH_4$ bands are shifted to wavelengths associated with $CH_4$ highly diluted in $N_2$ ice (Quirico and Schmitt 1997; Grundy et al. 2010), while Makemake's $CH_4$ bands are only slightly shifted, indicating that much less of its $CH_4$ is diluted in $N_2$ ice (e.g., Licandro et al. 2006a; Tegler et al. 2008). For Triton, we used the Quirico et al. (1999) "best" model, with all of the $CH_4$ simulated as being dissolved in $N_2$. For Makemake, we constructed a simple model consisting of 300 µm grains of undiluted $CH_4$ ice. These simple models were computed for a series of D/H ratios, shown in Fig. 7.

For a given D/H ratio, the $CH_3D$ bands in Fig. 7 are less deep in the Triton model than in the clean $CH_4$ ice model, because there is less $CH_4$ absorption overall. This is partly because mean optical path lengths in $CH_4$ ice are smaller on Triton and partly because 45% of the Triton model surface has no $CH_4$ ice on it at all, being composed of $H_2O$ and $CO_2$ ices. It is also due to the fact that, at longer wavelengths, continuum absorption (absorption at wavelengths without apparent absorption bands) is much higher in the diluted $CH_4$ optical constants of Quirico and Schmitt (1997) than it is in the pure $CH_4$ optical constants of Grundy et al. (2002), especially beyond 4 µm, where they differ by about two orders of magnitude. The much higher continuum absorption in the diluted optical constants has the effect of diminishing the spectral contrast of weak absorption bands in the reflectance model and diminishes the sensitivity of the 4.56 µm band so much in the Triton model that the 2.47 µm band is almost as sensitive for detecting $CH_3D$ ice. It is unclear whether this much higher continuum in the diluted ice is real, since absorption at continuum wavelengths is by its nature difficult to measure. Future experiments with thicker samples of $CH_4$ diluted in $N_2$ could perhaps resolve this question. It is also worth noting that the depths of observed $CH_3D$ absorption bands are highly dependent on the surface configuration, so it is essential to do a full spectral model to simultaneously constrain the D/H ratio along with other compositional and textural parameters. A number of additional factors liable to complicate interpretation of D/H ratios in solar system $CH_4$ ice are discussed by Brown and Cruikshank (1997), including possible vibrational coupling effects between isotopically shifted $CH_3D$ bands and the bands of ordinary $CH_4$. Isotopic stratification is also possible due to fractionation effects associated with volatile transport. Indeed, spatially resolved D/H observations coupled with compositional and geomorphological studies could be particularly revealing of volatile transport history, and with these new laboratory data, may soon become possible when New Horizons explores Pluto in 2015 (Young et al. 2008). Finally, D/H ratios in outer solar system $CH_4$ ice could tell a distinct story from D/H ratios in water, since water and methane have different chemical histories and different proton exchange behaviors.

## 4. Conclusion

Between 1 and 6 µm, we investigate spectral regions where strong $CH_3D$ ice absorptions coincide with weak absorption by ordinary $CH_4$ ice, a condition that could enable remote measurement of $CH_3D$ and thus D/H ratios in $CH_4$ ice even when $CH_3D$ is vastly outnumbered by ordinary $CH_4$ molecules. Three $CH_3D$ bands at 2.47, 2.87, and 4.56 µm look especially promising for this purpose. All three wavelengths can be observed from ground-based telescopes, albeit with some difficulty due to absorption by the terrestrial atmosphere. They would be readily accessible to space-based instruments, as would a fourth band at 4.34 µm. We report new temperature-dependent absorption coefficients for the spectral contribution of $CH_3D$ in $CH_4$ ice, subject to the assumption of no appreciable isotopic fractionation of $CH_4$ in our experiment between gas, liquid, and solid phases (if such fractionation did enrich the $CH_3D$ in our ice samples, we have overestimated its absorption coefficients). We also report temperature-dependent absorption coefficients for $CH_3D$ diluted in $\beta$ $N_2$ ice, further assuming that the integrated areas of the $CH_3D$ bands remain unchanged for $CH_3D$ in $CH_4$ and in $N_2$ ice. As with ordinary $CH_4$, $CH_3D$ absorption bands shift to slightly shorter wavelengths and become narrower when $CH_3D$ is dispersed in $N_2$ ice. We expect that these new laboratory measurements will enable the scientific potential of deuterium to be extended by observers to smaller bodies that formed further out in the protoplanetary nebula.

## Acknowledgments

We are grateful to the Mt. Cuba Astronomical Foundation for funding the hardware upgrades described in this paper. Grundy and Bovyn thank the National Aeronautics and Space Administration's Planetary Geology and Geophysics program for grant NNX10AG43G to Lowell Observatory that enabled their contributions to this work. Morrison thanks the National Science Foundation's Research Experience for Undergraduates grant AST-1004107 to Northern Arizona University that enabled her contribution. We are grateful to two anonymous reviewers for their many constructive suggestions to improve this manuscript. Finally, we thank the free and open source software communities for empowering us with key tools used to complete this project, notably Linux, the GNU tools, OpenOffice.org, MySQL, Evolution, and FVWM.